\documentclass[aps,prl,longbibliography,showpacs,superscriptaddress,twocolumn]{revtex4-1}%
\usepackage{bm,amsmath,amssymb}
\usepackage{amsfonts}
\usepackage[english]{babel}
\usepackage[colorlinks=true,linkcolor=blue,citecolor=blue,urlcolor=blue]{hyperref}

\usepackage{txfonts}

\usepackage{graphicx}

\begin{document}
\title{Hyper- and hybrid nonlocality}
\author{Yanna Li}
\thanks{These authors contributed equally to this work.}
\affiliation{Institute of Theoretical Physics and Department of
Physics,State Key Laboratory of Quantum Optics and Quantum Optics
Devices, Collaborative Innovation Center of Extreme Optics, Shanxi
University, Taiyuan 030006, China}
\author{Manuel Gessner}
\thanks{These authors contributed equally to this work.}
\affiliation{QSTAR, INO-CNR and LENS, Largo Enrico Fermi 2, I-50125 Firenze, Italy}
\author{Weidong Li}
\thanks{wdli@sxu.edu.cn}
\affiliation{Institute of Theoretical Physics and Department of Physics,State Key Laboratory of Quantum Optics and Quantum Optics
Devices, Collaborative
Innovation Center of Extreme Optics, Shanxi University, Taiyuan 030006, China}
\author{Augusto Smerzi}
\affiliation{QSTAR, INO-CNR and LENS, Largo Enrico Fermi 2, I-50125 Firenze, Italy}

\begin{abstract}
The controlled generation and identification of quantum correlations, usually encoded in either qubits or continuous degrees of freedom, builds the foundation of quantum information science. Recently, more sophisticated approaches, involving a combination of two distinct degrees of freedom have been proposed to improve on the traditional strategies. Hyperentanglement describes simultaneous entanglement in more than one distinct degree of freedom, whereas hybrid entanglement refers to entanglement shared between a discrete and a continuous degree of freedom. In this work we propose a scheme that allows to combine the two approaches, and to extend them to the strongest form of quantum correlations. Specifically, we show how two identical, initially separated particles can be manipulated to produce Bell nonlocality among their spins, among their momenta, as well as across their spins and momenta. We discuss possible experimental realizations with atomic and photonic systems.
\end{abstract}

\date{\today}
\pacs{03.65Ud, 03.67.Bg, 03.67.Mn, 03.75.Gg}
\maketitle

\textit{Introduction.}---Sharing quantum correlations between
distant parties is an indispensable condition for most tasks in
quantum communication \cite{NielsenChuang}. In the most common
scenario, quantum information is encoded into a single,
well-controlled degree of freedom (DOF), such as spin, polarization,
or external degrees of freedom \cite{RevModPhys.74.145,Weedbrook}.
In some cases, however, establishing entanglement among several DOF
can provide a decisive advantage
\cite{Kwiat97,PhysRevA.65.032317,PhysRevA.68.042313,Barreiro2008,PhysRevLett.105.170404,PhysRevA.85.032307,PhysRevA.87.022326,Takeda2013,Andersen2015}.
For example, so-called hyperentanglement, i.e., entanglement in
multiple DOF \cite{Kwiat97}, can improve the capacity of dense
coding in linear optics \cite{Barreiro2008}, or enhance the
performance of quantum teleportation \cite{PhysRevA.68.042313}.
Similarly, architectures using hybrid entanglement, i.e.,
entanglement across discrete and continuous variables
\cite{PhysRevA.65.032317,Gessner2016}, have been suggested as a
promising platform for quantum information, being able to overcome
the limitations posed by the finite detection efficiencies of
traditional approaches to quantum cryptography and computing
\cite{PhysRevLett.105.170404,Andersen2015}.

Quantum correlations can be classified in a hierarchical order \cite{Werner1989,PhysRevLett.98.140402,Adesso2016}; 
their strongest manifestations are Bell correlations, or nonlocality
\cite{RevModPhys.86.419}. Certain quantum information protocols,
such as the realization of secure quantum communication
\cite{Ekert1991}, explicitly require such Bell correlations,
rendering the mere presence of entanglement insufficient. Quantum
correlations involving hybrid variables of single particles are
routinely generated and detected in atom- and photon-based
experiments
\cite{Monroe1996,PhysRevLett.77.4887,Haroche2013,Wineland2013,Tang2014,Gessner2014NP}.
Recently, hybrid entanglement of photons in spatially or temporally
separated modes was achieved \cite{Jeong2014,Morin2014}. Experiments
with hyperentangled photon states have been reported
\cite{PhysRevLett.95.260501,PhysRevA.68.042313,Barreiro2008},
whereas these states contain no correlations across the different
DOF.

\begin{figure}[tb]
\centering
\includegraphics[width=0.4\textwidth]{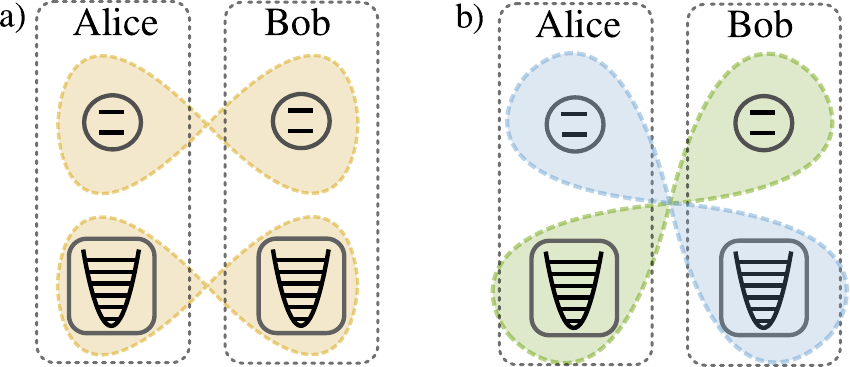}
\caption{Hyper- and hybrid nonlocality. (a) Hypernonlocality represents the simultaneous presence of Bell correlations among more than one DOF of two spatially separated particles. (b) Hybrid nonlocality identifies Bell correlations among the discrete DOF of one particle and the continuous DOF of another distant particle.}%
\label{fig:hyperhybrid}%
\end{figure}

Here, we propose a scheme to generate Bell correlations between
internal and external DOF of two spatially separated particles, as
well as across those two DOF, see Fig.~\ref{fig:hyperhybrid}. The
correlations are revealed through the violation of a series of
Clauser-Horne-Shimony-Holt (CHSH) inequalities \cite{CHSH}. The
fundamental element of our scheme is a hybrid beam splitter which
simultaneously entangles internal and external DOF. We show how such
an operation may be experimentally realized with atomic and quantum
optical systems. This allows us to explore new possibilities for the
design of efficient quantum information protocols that make use of
Bell correlations across and within several DOF at the same time.

\textit{Hybrid beam splitters.}---We consider particles with
internal (e.g., spin) and external (e.g., momentum) DOF.
Correlations between the two degrees of freedom can be induced by
processes of the type
\begin{align}\label{eq:hybrid}
\vert\downarrow,\mathbf{p}_{\downarrow}\rangle\longrightarrow\alpha \vert\downarrow,\mathbf{p}_{\downarrow}\rangle+\beta\vert\uparrow,\mathbf{p}_{\uparrow}\rangle,
\end{align}
where $\vert i,\mathbf{p}\rangle$ describes a particle with spin state $\vert i\rangle$ and momentum $\mathbf{p}$, and $|\alpha|^2+|\beta|^2=1$. Such processes are encountered in many physical systems, including cavity-QED systems \cite{Haroche2013}, birefringent optical materials \cite{Tang2014}, and trapped ions under sideband transitions \cite{Wineland2013,Gessner2014NP}. In the following we focus on identical bosonic particles, where the above process is combined with an interference effect due to the indistinguishability of the particles \cite{PhysRevLett.59.2044,Lopes2015}. As we will discuss later in further detail, such an effect can be generated with the aid of a two-photon Raman process \cite{Aspect1989,PhysRevLett.67.181,PhysRevA.45.342,Borde1997}, or by combining quarter-wave plates with polarizing beam splitters in linear optics \cite{WZ01,Kok07}.

We now turn to a more convenient second-quantized description, with
bosonic operators $a_{i,\mathbf{p}}$ with $\vert
i,\mathbf{p}\rangle=a^{\dagger}_{i,\mathbf{p}}\vert 0\rangle$, where
$\vert 0\rangle$ is the vacuum. These operators satisfy the
canonical commutation relations:
\begin{align}
[a_{i,\mathbf{p}_i},a_{j,\mathbf{p}_j}]=0,\quad[a_{i,\mathbf{p}_i},a^{\dagger}_{j,\mathbf{p}_j}]=\delta(\mathbf{p}_i-\mathbf{p}_j)\delta_{ij}.
\end{align}
Let us now consider the process Eq.~(\ref{eq:hybrid}) for
$|\alpha|^2=|\beta|^2$, restricting to two external orthogonal
modes, described by $a_{j,\mathrm{in}_{1,2}}$  and
$a_{j,\mathrm{out}_{1,2}}$, respectively. Based on the process
Eq.~(\ref{eq:hybrid}), a balanced two-mode beam splitter in both
internal and external DOF can be described as
\begin{subequations}\label{eq:hbs}
\begin{align}\label{eq:hbs1}
\begin{pmatrix}
a_{\downarrow,\mathrm{out}_1}\\
a_{\uparrow,\mathrm{out}_2}
\end{pmatrix}=
\frac{1}{\sqrt{2}}
\begin{pmatrix}
1 & i\\
i &  1
\end{pmatrix}
\begin{pmatrix}
a_{\downarrow,\mathrm{in}_1} \\
a_{\uparrow,\mathrm{in}_2}
\end{pmatrix}
\end{align}
and
\begin{align}\label{eq:hbs2}
\begin{pmatrix}
a_{\downarrow,\mathrm{out}_2}\\
a_{\uparrow,\mathrm{out}_1}\\
\end{pmatrix}=
\frac{1}{\sqrt{2}}
\begin{pmatrix}
1 & i \\
i & 1 \\
\end{pmatrix}
\begin{pmatrix}
a_{\downarrow,\mathrm{in}_2} \\
a_{\uparrow,\mathrm{in}_1} \\
\end{pmatrix}.
\end{align}
\end{subequations}
This hybrid beam splitter leads to the generation of spin-momentum
correlations.

\begin{figure}[bt]
\centering \vspace{0.in}
\includegraphics[width=0.49\textwidth]{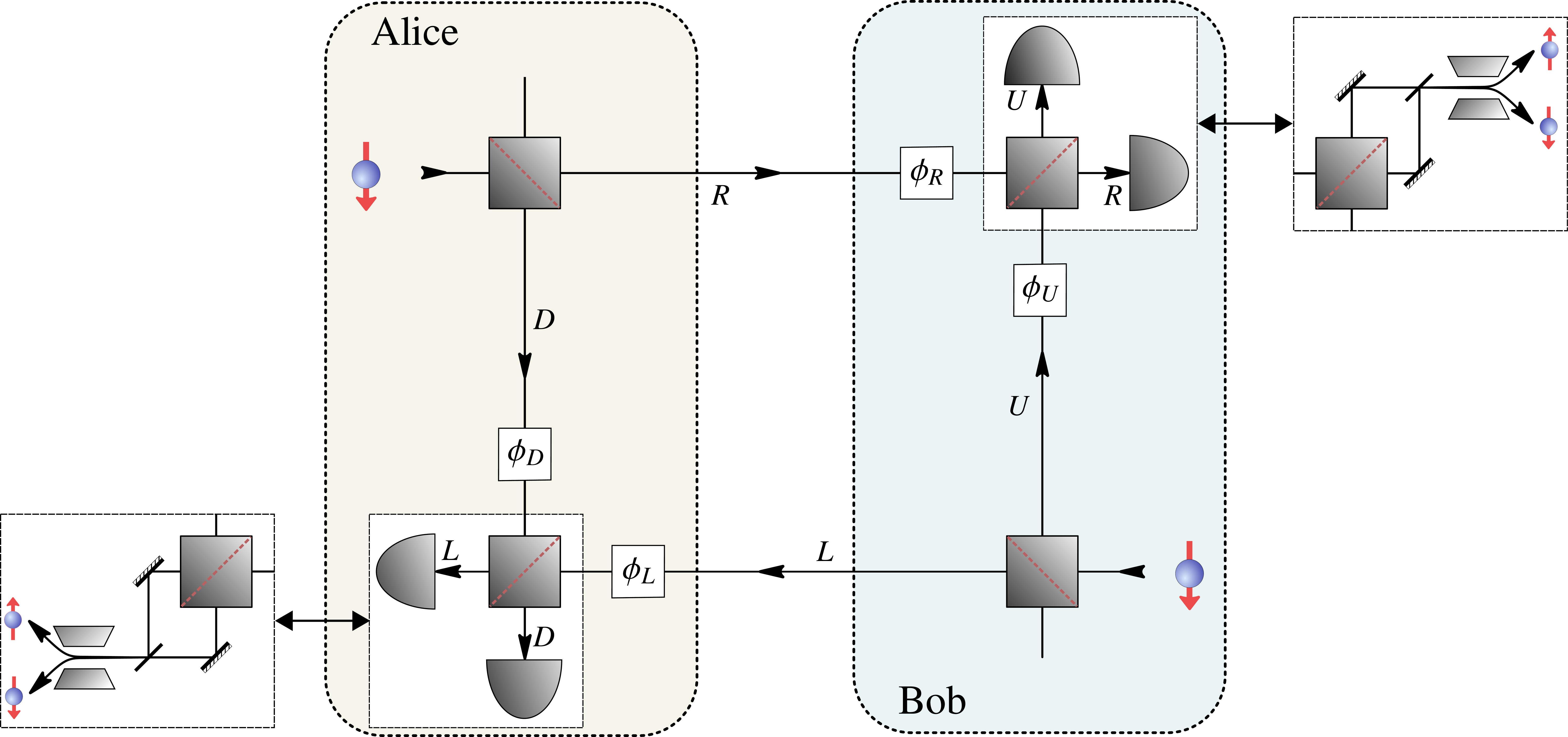}
\caption{Experimental scheme for the generation and verification of intra and inter-DOF nonlocality. Alice and Bob both prepare one particle in a spin-$\left\vert\downarrow\right\rangle$ state and submit it to a hybrid beam splitter. One of the output ports is sent to their local laboratory while the other is sent to the opposite party. By mixing the local and the received copy using a second hybrid beam splitter, the desired correlations are established. Both parties now measure either spin or external DOF of their received particles, as depicted by the interchangeable measurement devices (white boxes). The recorded data from the events in which both parties receive exactly one particle violate a suitable CHSH inequality, independently of the measured DOF.}
\label{fig:scheme}%
\end{figure}

\textit{Generation of intra- and inter-DOF nonlocality.}---Let us
now consider the setup depicted in Fig.~\ref{fig:scheme}. Such an
array was first proposed by Yurke and Stoler \cite{PhysRevA.46.2229}
who -- using a single DOF and conventional beam splitters --
demonstrated that Bell correlations can be generated with identical
particles from independent sources \cite{PhysRevLett.68.1251}; see
Refs. \cite{Halder2007,Neder2007} for experimental realizations with
photons and electrons. Our envisioned sequence is represented in
form of an array of hybrid beam splitters and phase shifts,
involving four orthogonal external modes $R$, $L$, $U$, and $D$.
Particles that exit in the modes $L$ and $D$ are received by Alice,
whereas particles in modes $R$ and $U$ are sent to Bob's detectors.
Alice further controls the phases $\phi_L$ and $\phi_D$ while Bob
has access to the phases $\phi_R$ and $\phi_U$.

We consider two particles entering the setup in the modes $R$ and
$L$, both with spin state $\left\vert\downarrow\right\rangle$; i.e.,
the initial state is
$\vert\Psi_0\rangle=a^{\dagger}_{\downarrow,R}a^{\dagger}_{\downarrow,L}\vert0\rangle$.
We may consider these particles created within the setups of Alice
and Bob. After sending their particle into a hybrid beam splitter,
one of the output ports ($R$ or $L$) is sent to the respective other
party, while the other one ($U$ or $D$) remains locally accessible
(see Fig.~\ref{fig:scheme}). Next, a state-dependent phase shift is
imprinted. While in general one may consider also spin-dependent
phase shifts, for our purposes it suffices to employ phase shifts
that depend only on the external modes:
$a_{j,\mathrm{out}}=e^{i\phi_{\mathrm{in}}}a_{j,\mathrm{in}}$.
Finally, employing a second hybrid beam splitter, the local mode is
mixed with the mode that receives
the other parties' particle, followed by a measurement. The measurement can be either of the external modes, without measuring the spin state, or of the spin state, without discriminating between external modes. The described combination of two pairs of hybrid beam splitters and path-dependent phase shifts (Fig.~\ref{fig:scheme}) transforms the initial state $\vert\Psi_0\rangle$ into
\begin{align}
|\Psi\rangle&=\frac{1}{4}[e^{i\phi_R}(a_{\downarrow,R}^{\dagger}
+ia_{\uparrow,U}^{\dagger})+ie^{i\phi_D}(a_{\uparrow,D}^{\dagger}
+ia_{\downarrow,L}^{\dagger})]\notag\\
&\quad\times[e^{i\phi_L}
(a_{\downarrow,L}^{\dagger}+ia_{\uparrow,D}^{\dagger})+ie^{i\phi_U}
(a_{\uparrow,U}^{\dagger} +ia_{\downarrow,R}^{\dagger})]|0\rangle.
\end{align}

\textit{Violation of CHSH inequalities.}---First, we consider
coincidence measurements only of the external DOF. The detection
probabilities for events where both Alice and Bob each receive
exactly one particle are given by
\begin{align}\label{eq:exttab}
\begin{array}{l|cc}
 & B:R & B:U \\\hline
A:D & \frac{1}{4}\sin^2 \phi & \frac{1}{4}\cos^2 \phi\\
A:L & \frac{1}{4}\cos^2 \phi & \frac{1}{4}\sin^2 \phi
\end{array},
\end{align}
as a function of the total phase shift
$\phi=(\phi_D-\phi_L-\phi_R+\phi_U)/2$. They coincide with those of
the linear optical scheme based only on a single DOF as considered
by Yurke and Stoler \cite{PhysRevA.46.2229}. As was shown in their
work, one may use these events to define dichotomic variables as a
function of the observed output port. Specifically, assigning the
event $+1$ to clicks in the respective upper detector ($L$ for Alice
and $U$ for Bob) and $-1$ to clicks in the respective lower detector
($D$ for Alice and $R$ for Bob), we obtain from
Eq.~(\ref{eq:exttab}) the probabilities $P_{ij}$ for coincidence
events of Alice observing $i=\pm 1$ and Bob observing $j=\pm 1$. The
normalized expectation value
\begin{align}
E(\phi_A,\phi_B)=\frac{P_{++}-P_{-+}-P_{+-}+P_{--}}{P_{++}+P_{-+}+P_{+-}+P_{--}}=-\cos (\phi_A-\phi_B),
\end{align}
is a function of the two relative phases $\phi_A=\phi_D-\phi_L$ and
$\phi_B=\phi_U-\phi_R$, which are under the local control of Alice
and Bob, respectively. Introducing two detector settings for each
party, i.e., angles $\phi_A^{0},\phi_B^{0},\phi_A^{1},\phi_B^{1}$,
we can formulate the CHSH inequality \cite{CHSH},
\begin{align}
|E(\phi_A^{0},\phi_B^{0})+E(\phi_A^{1},\phi_B^{0})+E(\phi_A^{0},\phi_B^{1})-E(\phi_A^{1},\phi_B^{1})|\leq 2,
\end{align}
whose violation implies that the recorded events are incompatible with local realism \cite{EPR1935,Bell1964}. For $\phi_A^{0}=0$, $\phi_A^{1}=\pi$, $\phi_B^{0}=\pi/4$, and $\phi_B^{1}=-\pi/4$ we obtain the maximal violation of the inequality permitted by quantum mechanics, i.e., Tsirelson's bound $2\sqrt{2}>2$ \cite{Cirel'son1980}.

Rather than measuring the external DOF, i.e., the particles' output
port, Alice and Bob can instead choose to measure the received
particles' spin states. For the events where Alice and Bob
coincidentally receive exactly one particle, the probabilities for
spin measurements are given by
\begin{align}
\begin{array}{l|cc}
 & B:\downarrow & B:\uparrow \\\hline
A:\downarrow & \frac{1}{4}\cos^2 \phi & \frac{1}{4}\sin^2 \phi \\
A:\uparrow & \frac{1}{4}\sin^2 \phi & \frac{1}{4}\cos^2 \phi
\end{array}.
\end{align}
Here $A$ ($B$) represents a particle received by Alice (Bob), i.e., exiting the output ports $D$ or $L$ ($R$ or $U$). Assigning the value $+ 1$ to the detection event of $\left\vert\uparrow\right\rangle$ and $-1$ to $\left\vert\downarrow\right\rangle$, we obtain the expectation value $E(\phi_A,\phi_B)=\cos(\phi_A-\phi_B)$. This produces the same violation of the CHSH inequality as before, this time, however, by measuring only spin variables.

Finally, we consider hybrid detection events. In this scenario, Alice records which of the two output detectors click without registering the spin state, while Bob records only the spin state of the particle that exits on his side in coincidence, regardless of the output port (or vice versa). The combined events are described by the probabilities
\begin{align}
\begin{array}{l|cc}
 & B:R & B:U \\\hline
A:\downarrow & \frac{1}{4}\cos^2 \phi & \frac{1}{4}\sin^2\phi \\
A:\uparrow & \frac{1}{4}\sin^2 \phi & \frac{1}{4}\cos^2  \phi
\end{array}
\quad\text{or}\quad
\begin{array}{l|cc}
 & B:\downarrow & B:\uparrow \\\hline
A:D & \frac{1}{4}\sin^2 \phi & \frac{1}{4}\cos^2 \phi \\
A:L & \frac{1}{4}\cos^2 \phi & \frac{1}{4}\sin^2 \phi
\end{array}.
\end{align}
Assigning again the events $\pm 1$ to the spin or external measurement results as above, we obtain $E(\phi_A,\phi_B)=\pm\cos(\phi_A-\phi_B)$, and consequently the violation of the CHSH inequality by means of hybrid measurements.

To summarize, the state generated by the array of hybrid beam
splitters describes two particles with nonlocal Bell correlations
among their spins, their external DOF, as well as hybrid nonlocality
across the two DOF; see Fig.~\ref{fig:hyperhybrid}. Nonlocality is
revealed regardless of whether Alice or Bob, independently of each
other, chooses to perform spin or external measurements.

Central to the generation of these correlations is the hybrid beam
splitter Eq.~(\ref{eq:hbs}) which entangles the path of an incoming
particle with its internal state. Ultimately, when the particles
reach the detectors, this inter-DOF entanglement renders the choice
of DOF for the measurement irrelevant. The decisive role is played
by the first pair of hybrid beam splitters. Since the measurement is
always limited to a single DOF, the hybrid beam splitters employed
just before the measurement can be replaced by single-DOF beam
splitters in the variable that is subsequently measured, without
affecting the detection probabilities.

The generated quantum correlations are entirely due to the symmetrization of the bosonic two-particle wave function \cite{Wasak16}. The present scheme thus describes a possible way to use these correlations, effectively transferring them from the inaccessible particle labels to the distant modes of Alice and Bob.

\begin{figure}[tb]
\centering
\includegraphics[width=0.49\textwidth]{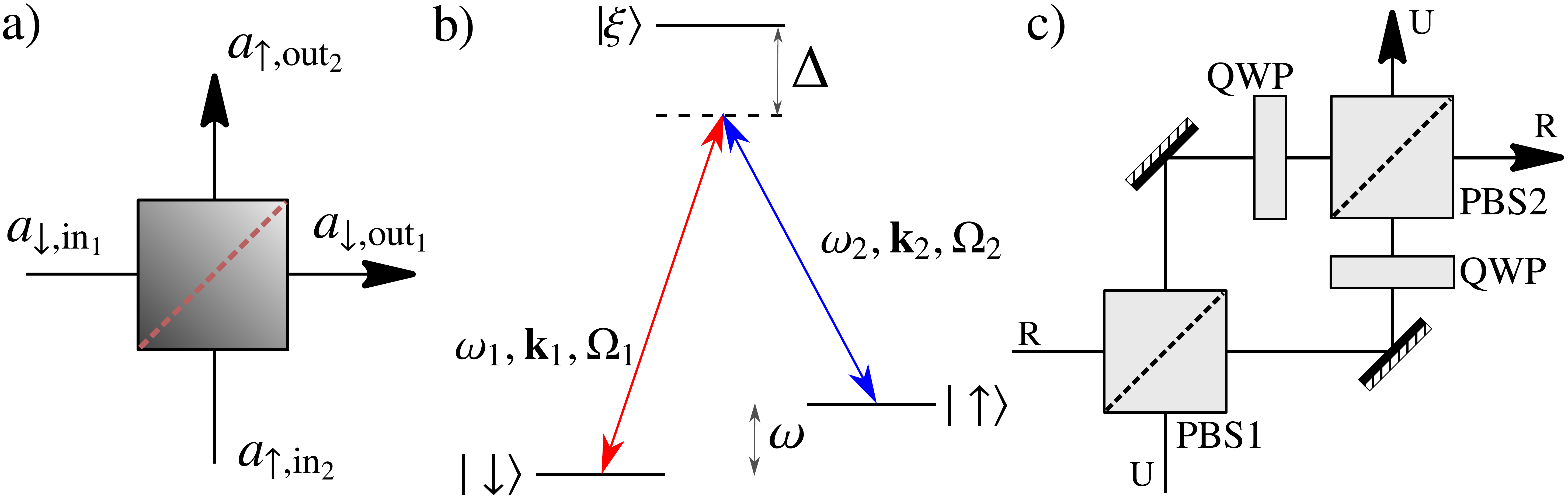}
\caption{Hybrid beam splitters. (a) Hybrid beam splitters combine deflections with a change of the internal quantum state. (b) The two-photon stimulated Raman transition couples the states $\left\vert\downarrow\right\rangle$ and $\left\vert\uparrow\right\rangle$, whose energy difference $\omega=\omega_1-\omega_2$ is resonant with the two light fields with an effective Rabi frequency of $\Omega=\Omega_1\Omega_2/(2\Delta)$ in the limit of $\Delta\gg\Omega_{1,2}$. In the process a momentum of $\hbar\mathbf{k}=\hbar(\mathbf{k}_1-\mathbf{k}_2)$ is transferred to the atom. The scheme realizes a pairwise hybrid beam splitter coupling with external modes $(\mathrm{in/out})_1=\mathbf{p}$ and $(\mathrm{in/out})_2=\mathbf{p}+\hbar\mathbf{k}$. (c) Optical realization of a hybrid beam splitter with external modes $(\mathrm{in/out})_1$=$R$ and $(\mathrm{in/out})_2$=$U$, and internal polarization states. The first polarizing beam splitter (PBS1) transmits all photons in one polarization state, while deflecting all others. After passing through quarter-wave plates (QWP) the photons enter PBS2, whose transmission properties are opposite to PBS1. The combination realizes the hybrid coupling, as described in Eq.~(\ref{eq:hbs}).}%
\label{fig:Raman}%
\end{figure}

\textit{Realization of hybrid beam splitters.}---An experimental
realization of the hybrid beam splitter can be achieved with the aid
of a Raman process. An atom in the presence of bichromatic laser
light can undergo a two-photon process from one internal state to
another,
$\left\vert\downarrow\right\rangle\longrightarrow\left\vert\uparrow\right\rangle$
(Fig.~\ref{fig:Raman}). In that process the atom absorbs one photon
with momentum $\hbar\mathbf{k}_1$ and reemits another with momentum
$\hbar\mathbf{k}_2$, thereby acquiring a total momentum shift of
$\hbar\mathbf{k}=\hbar(\mathbf{k}_{1}-\mathbf{k}_{2})$. This
stimulated Raman transition has been widely exploited in atom
cooling
\cite{Aspect1989,PhysRevA.45.342,PhysRevLett.69.1741,Wineland98},
matter-wave interferometers
\cite{PhysRevLett.67.181,Weiss94,Borde1997,RevModPhys.81.1051,Rosi2014},
and the generation of synthetic spin-orbit couplings
\cite{Zhai2015,Goldman2014}. Most importantly, the internal state of
the atom becomes correlated with its momentum, producing a coherent
superposition of $\left\vert\downarrow,\mathbf{p}\right\rangle $ and
$\left\vert\uparrow,\mathbf{p}+\hbar\mathbf{k}\right\rangle$. The
bias of the superposition can be experimentally controlled by
adjusting the interaction time $\tau$ of the Raman process. In
particular, a balanced two-mode process can be realized with a
$\pi/2$ pulse by setting $\tau=\pi/(2\Omega)$, where $\Omega$ is the
effective two-photon Rabi frequency (Fig.~\ref{fig:Raman}). This
leads to the transformation
$\vert\downarrow,\mathbf{p}\rangle\longrightarrow(\vert
\downarrow,\mathbf{p}\rangle +e^{i\phi}\vert
\uparrow,\mathbf{p}+\hbar\mathbf{k}\rangle)/\sqrt{2}$ and $\vert
\uparrow,\mathbf{p}+\hbar\mathbf{k}\rangle\longrightarrow(\vert\uparrow,\mathbf{p}+\hbar\mathbf{k}\rangle+e^{i\phi}\vert
\downarrow,\mathbf{p}\rangle)/\sqrt{2}$, where the accumulated phase
$\phi$ can be controlled adjusting the phases of the lasers. The
transformation has indeed the form of Eq.~(\ref{eq:hybrid}) and
realizes the required hybrid coupling between the resonant pair of
states $\vert\downarrow,\mathbf{p}\rangle$ and $\vert
\uparrow,\mathbf{p}+\hbar\mathbf{k}\rangle$. Manipulating
$\mathbf{k}$ allows us to select the pair of states which is
coherently coupled by the process. For instance, by changing the
sign of $\mathbf{k}$ (i.e., the orientation of the two lasers), the
pair $\vert\downarrow,\mathbf{p}+\hbar\mathbf{k}\rangle$ and
$\vert\uparrow,\mathbf{p}\rangle$ is coupled. We emphasize that each
of the hybrid beam splitters in Fig.~\ref{fig:scheme} effectively
only acts on a single pair of states, due to the choice of the
initial state. Hence, it suffices to realize either
Eq.~(\ref{eq:hbs1}) or Eq.~(\ref{eq:hbs2}) for a suitable set of
external states by means of the Raman process. Detailed treatments
of the Raman process can be found in the literature; see, e.g.,
Refs.~\cite{Aspect1989,PhysRevA.45.342,PhysRevLett.67.181,Borde1997,Weiss94,Wineland98,RevModPhys.81.1051}.
Spin and momentum states can further be manipulated individually
with high accuracy using resonant laser manipulations and Bragg
techniques, respectively
\cite{PhysRevLett.67.181,PhysRevLett.114.063002,PhysRevLett.82.871,RevModPhys.81.1051}.

We remark that even if the momenta of the two coupled modes are not
perpendicular in $\mathbb{R}^3$, as is commonly the case in atom
interferometry experiments, the process still realizes a hybrid beam
splitter. The only important aspect is orthogonality in Hilbert
space, which is always achieved by the Raman process for nonzero
momentum transfer.

A hybrid beam splitter may also be realized in optical systems,
creating correlations between the photons' polarization (internal)
and their path (external). A combination of two polarizing beam
splitters and quarter-wave plates, as depicted in
Fig.~(\ref{fig:Raman}c), realizes the required coupling described by
Eq.~(\ref{eq:hbs}) with external modes $(\mathrm{in/out})_1$=$R$ and
$(\mathrm{in/out})_2$=$U$, and $\downarrow,\uparrow$ correspond to
horizontal and vertical polarization states, respectively. The
polarizing beam splitter PBS1 transmits all photons in a specific
polarization state, e.g., $\downarrow$, while deflecting photons
with an orthogonal polarization, e.g., $\uparrow$ (independently of
their incoming path) \cite{WZ01,Kok07}. The quarter-wave plates act
as beam splitters on the polarization state without affecting the
path, e.g., $a_{\uparrow,R}\rightarrow
(a_{\uparrow,R}+ia_{\downarrow,R})/\sqrt{2}$. The second polarizing
beam splitter PBS2 transmits the polarization state that was
deflected by PBS1 and deflects the previously transmitted one. The
described combination realizes a hybrid beam splitter in a linear
optical system.

Hybrid entanglement implies quantum correlations involving a
continuous variable. In the setup depicted in Fig.~\ref{fig:scheme},
the external DOF is limited to four possible states, which
effectively renders it discrete. Even though the scheme presented
here involves only a finite number of momentum states, the
underlying Hilbert space of the external DOF is unbounded and is
described by continuous variables \cite{Cavalcanti,Salles,Qian}. In
atomic systems, the parameter $\mathbf{k}$ of the Raman process can
be continuously tuned to generate quantum correlations between spins
and momenta of a continuum of possible state pairs. In quantum
optical systems, continuous variables in the form of quadratures can
be measured with homodyne techniques. The scheme in
Fig.~\ref{fig:scheme} can be extended by sharing a local oscillator
mode as common phase reference among Alice and Bob for quadrature
measurements \cite{Banaszek}.

\textit{Complete Bell-state analysis.}---Distinguishing between the
four maximally entangled Bell states,
$\left\vert\Psi_{\pm}\right\rangle =
(\left\vert\uparrow\downarrow\right\rangle\pm\left\vert\downarrow\uparrow\right\rangle)/\sqrt{2}$
and $\left\vert\Phi_{\pm}\right\rangle =
(\left\vert\downarrow\downarrow\right\rangle\pm\left\vert\uparrow\uparrow\right\rangle)/\sqrt{2}$,
is a fundamental ingredient for quantum information protocols
including quantum teleportation \cite{Bennett1993} and dense coding
\cite{Bennett1992}. The performance of these protocols depends
crucially on the number of Bell states that can be discriminated. To
identify a Bell state in optical experiments, Bell-state analyzers
based on the Hong-Ou-Mandel effect \cite{PhysRevLett.59.2044} are
employed \cite{PhysRevA.59.3295,Weinfurter1994,PhysRevA.51.R1727}.
The two potentially spin-entangled particles are distributed among
the two input ports of a standard beam splitter. The detection
events after the beam splitter can be unambiguously traced back to
specific Bell states among the two input modes. However, in some
cases the result may be inconclusive, and only two out of four Bell
states can be distinguished
\cite{PhysRevA.51.R1727,PhysRevA.59.116,PhysRevA.59.3295,CL01,URen03,KG03,Grice11}.
It is possible to circumvent this limitation by making use of
hyperentangled input states \cite{PhysRevA.68.042313}. Even without
involving hyperentangled states, we will see below that a hybrid
beam splitter can render a Bell-state analyzer susceptible to the
two Bell states that cannot be distinguished by the conventional
setup.

First, we recall the results \cite{PhysRevA.51.R1727} for a standard beam splitter that acts only on the external degrees of freedom and realizes the transformation
\begin{align}
\begin{pmatrix}
a_{j,\mathrm{out}_1}\\
a_{j,\mathrm{out}_2}
\end{pmatrix}
=\frac{1}{\sqrt{2}}\begin{pmatrix}
1 & i\\ i & 1
\end{pmatrix}
\begin{pmatrix}
a_{j,\mathrm{in}_1}\\
a_{j,\mathrm{in}_2}
\end{pmatrix},
\end{align}
for $j=\uparrow,\downarrow$. Exactly one particle is submitted into each of the two input ports $a_{j,\mathrm{in}_1}$ and $a_{k,\mathrm{in}_2}$. Output events with exactly one particle in each output port $a_{j,\mathrm{out}_1}$ and $a_{k,\mathrm{out}_2}$ with different spin states $j\neq k$ then identify the Bell state $\vert\Psi_{-}\rangle\simeq (a^{\dagger}_{\uparrow,\mathrm{in}_1}a^{\dagger}_{\downarrow,\mathrm{in}_2}-a^{\dagger}_{\downarrow,\mathrm{in}_1}a^{\dagger}_{\uparrow,\mathrm{in}_2})\vert 0\rangle\simeq a^{\dagger}_{\uparrow,\mathrm{out}_1}a^{\dagger}_{\downarrow,\mathrm{out}_2}\vert 0\rangle\simeq a^{\dagger}_{\downarrow,\mathrm{out}_1}a^{\dagger}_{\uparrow,\mathrm{out}_2}\vert 0\rangle$. Here, we use the symbol $\simeq$ to indicate equality up to a constant factor after disregarding events with more than one particle in each of the input ports. If instead both particles are registered in the same output port, again with different spin states, we reveal the Bell state $\vert\Psi_{+}\rangle\simeq (a^{\dagger}_{\uparrow,\mathrm{in}_1}a^{\dagger}_{\downarrow,\mathrm{in}_2}+a^{\dagger}_{\downarrow,\mathrm{in}_1}a^{\dagger}_{\uparrow,\mathrm{in}_2})\vert 0\rangle\simeq a^{\dagger}_{\uparrow,\mathrm{out}_1}a^{\dagger}_{\downarrow,\mathrm{out}_1}\vert 0\rangle\simeq a^{\dagger}_{\uparrow,\mathrm{out}_2}a^{\dagger}_{\downarrow,\mathrm{out}_2}\vert 0\rangle$. The two remaining Bell states $\vert\Phi_{\pm}\rangle$ cannot be unambiguously identified using this scheme.

Let us now consider a Bell-state analyzer based on the hybrid beam
splitter Eq.~(\ref{eq:hbs}). The detection events of two particles
exiting at different ports in the same spin state identify the Bell
state $\vert \Phi_-\rangle\simeq
(a^{\dagger}_{\downarrow,\mathrm{in}_1}a^{\dagger}_{\downarrow,\mathrm{in}_2}-a^{\dagger}_{\uparrow,\mathrm{in}_1}a^{\dagger}_{\uparrow,\mathrm{in}_2})\vert
0\rangle\simeq
a^{\dagger}_{\uparrow,\mathrm{out}_1}a^{\dagger}_{\uparrow,\mathrm{out}_2}\vert
0\rangle\simeq
a^{\dagger}_{\downarrow,\mathrm{out}_1}a^{\dagger}_{\downarrow,\mathrm{out}_2}\vert
0\rangle$. Conversely, events with two particles in the same output
port but with different spin states indicate the Bell state $\vert
\Phi_+\rangle\simeq
(a^{\dagger}_{\downarrow,\mathrm{in}_1}a^{\dagger}_{\downarrow,\mathrm{in}_2}+a^{\dagger}_{\uparrow,\mathrm{in}_1}a^{\dagger}_{\uparrow,\mathrm{in}_2})\vert
0\rangle\simeq
a^{\dagger}_{\uparrow,\mathrm{out}_1}a^{\dagger}_{\downarrow,\mathrm{out}_1}\vert
0\rangle\simeq
a^{\dagger}_{\uparrow,\mathrm{out}_2}a^{\dagger}_{\downarrow,\mathrm{out}_2}\vert
0\rangle$. The other two Bell states $|\Psi_{\pm}\rangle$ cannot be
unambiguously distinguished using the hybrid beam splitter. The
scheme is able to detect the two Bell states which remain unresolved
by the standard approach and vice versa. Hence, the two methods
complement each other and together provide sufficient means to
discriminate among all four Bell states.

\textit{Conclusions.}---Hybrid beam splitters mix internal and
external states of the incoming particles and thereby generate
hybrid entanglement across two DOF. Such processes can be realized
with existing optical and atomic systems. We demonstrated how this
effect can be exploited in a suitable array of beam splitters to
generate Bell correlations among multiple DOF of two independent
bosonic particles. The correlations are analyzed by means of CHSH
inequalities which are violated regardless of whether spin or
external measurements are performed on either of the particles.
First, this indicates hypernonlocality, i.e., nonlocality in more
than one DOF. Second, since nonlocality is also revealed when two
different DOF are measured, we further observe hybrid nonlocality,
i.e., Bell correlations between a discrete (spin) and a continuous
(momentum) DOF of spatially separated particles. Since nonlocality
represents the strongest form of quantum correlations, these can be
used to realize the most exigent quantum information protocols. By
efficiently harnessing hybrid nonlocal and/or hypernonlocal quantum
states, opportunities for the design of new protocols or for
improvements of existing schemes may open up.

\textit{Acknowledgments.}---This work was supported by the National Key R$\And$D Program of
China (No. 2017YFA0304500 and No. 2017YFA0304203), National Natural
Science Foundation of China (Grant No. 11374197), PCSIRT (Grant No.
IRT13076), the Hundred Talent Program of the Shanxi Province, and
the Program of State Key Laboratory of Quantum Optics and Quantum
Optics Devices (No:KF201703). M.G. acknowledges support by the
Alexander von Humboldt foundation.

\end{document}